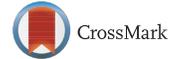

# Research Article
# A Genetic Framework Model for Self-adaptive Software

Enas Nafar and Said Ghoul

Research Laboratory on Bio-Inspired Systems Modeling, Faculty of Information Technology, Philadelphia University, Amman, Jordan

## Abstract

**Background:** Self-adaptive software changes its behavior at runtime without affecting the running system. It has recently been a rich research area. Lots of organizations have adopted it in their environments to accommodate with changing requirements. Lots of bio-inspired research works, which are better than the conventional ones have been conducted in the area of self-adaptive software. All of them have focused on the external behavior of biological entities (like birds, ants, immunity, etc.) without going in depth into their genetic material that causes this behavior and constitutes the challenge the work presented in this study dealt with. **Materials and Methods:** This study proposes a solution to the above current challenge by developing a framework model for self-adaptive software; inspired by the adaptation (evolution) of biological entities and taking into consideration the role of genetic material in the adaptation process. Its scope is limited to changes that take place at runtime but that are known at design. **Results:** The obtained framework model was evaluated through its reuse in software objects evolution. The practical and theoretical obtained results were valuable in the object-oriented paradigm. The proposed framework completes the others bio-inspired research current works by providing a natural implementing way. The integration of the current bio-inspired approaches (which deal with natural entities behaviors external modeling) with the proposed framework (which deals with genetics-inspired internal modeling of these behaviors) will lead to homogenous and coherent bio-inspired approaches to self-adaptive software. **Conclusion:** The proposed framework is limited to self-adaptations predicted at the requirements and design steps in self-adaptive software engineering, which is significant in practice. However, the unpredicted adaptation (to unpredicted errors, environment requirements, etc.) will be a genetics-inspired approach real challenge. Separate evaluation of the proposed framework performance is not determinant. However, the performance evaluation of the actual bio-inspired hybrid approaches against the proposed integrated ones (which is impossible to achieve actually) will be valuable. It might be expected that the integrated ones will be better (in the whole self-adaptive software engineering processes) than the hybrid current ones. The homogeneity of approaches has its important impact.

**Key word:** Self-adaptive software, bio-inspired self-adaptive software, genetics-inspired software modeling

## INTRODUCTION

An essential phase of software development lifecycle is software evolution (maintenance). It has been commonly accepted that software, which implements real world applications, must continually evolve. If software does not evolve, it will not fulfill the continuously changing requirements and thus, it will become outdated earlier than expected[1,2].

Software evolution is usually performed during scheduled down-times of the system; compromising the system's availability. Thus, the whole maintenance process is mainly performed off-line guided by human-driven change management activities and decoupled from the running system[2]. To deal with changes that take place at runtime, without affecting system's availability; software engineers have turned to self-adaptive software. This kind of software is capable of evaluating and changing its own behavior, whenever the evaluation shows that the software is not accomplishing what it was intended to do or when better functionality or performance may be possible[3]. The change is done by adjusting attributes (parameters) or artifacts of the system in response to changes in the system itself or in its environment[1].

There have been different proposed approaches to deal with self-adaptive software[4]. Some of them are based on conventional techniques such as Agents[5,6], Petri Nets[7] and UML[8]. Whereas some others are based on bio-inspired techniques[9,10] which are emergent and promising now-a-days[11]. So far, all researches concerned with bio-inspired self-adaptive software, have based their work on imitating the external behavior of biological entities like neural networks[12], cells[13], immunology[14] and ant colony[15] rather than taking advantage of the internal capabilities that exit within those living entities, like genetic material[16].

This study proposes a solution to this insufficiency. It proposes a genetics-inspired framework for self-adaptive software; inspired by the adaptation (evolution) of biological entities and taking into consideration the role of genes in this process. Consequently, it complete the others bio-inspired research works by providing a natural implementing way. The integration of the current bio-inspired approaches (which deal with the natural entities behaviors external modeling) with the proposed framework (which deals with genetics-inspired internal modeling of these behaviors) will lead to homogenous and coherent bio-inspired approaches to self-adaptive software. Without this integration the current bio-inspiration remains hybrid of natural inspiration and artificial (with computer paradigms) internal representation and consequently not homogenous.

A first evaluation was by reusing this framework in software object evolution[17]. The conceptual and practical obtained results were valuable in object-oriented paradigm. However, this separate evaluation of is not sufficient. The performance evaluation the actual bio-inspired hybrid approaches against the proposed integrated ones must be evaluated. Unfortunately, it is impossible to achieve currently this evaluation because the integrated approaches are just what this study proposes; but it might be expected that the integrated ones will be better (in the whole self-adaptive software engineering processes) than the hybrid current ones. The homogeneity of approaches has its important impact.

## MATERIALS AND METHODS

Self-adaptive software changes its behavior at runtime without affecting the running software. Adaptation of software usually deals with its features and behaviors. This study is concerned with adaptation that is pre-planned in a genetics-inspired framework that defines software lifecycle (Fig. 1). The following terminology, inspired from natural genetics will be used throughout this study.

**Business software database:** A set of all possible features concerning a certain business domain (similar to genome concept in genetics).

**A software configuration:** A selected subset of compatible features in a business software database; composing a release when being executed (similar to genotype concept in genetics).

**A software instance:** An operational software (release) of a certain software configuration (similar to phenotype concept in genetics).

**Genetic framework adaptation concepts:** This study introduces some key generic concepts of the genetic adaptation framework, which will be used throughout this study.

**Genetic adaptation program (GAProg):** It is a program framework that specifies what features are needed for a certain adaptation (Fig. 2). This framework is specified as it follows:

```
GAProg <ID>
{ Enable (<feature>,)+;
 Disable (<feature>,)+;
}
```

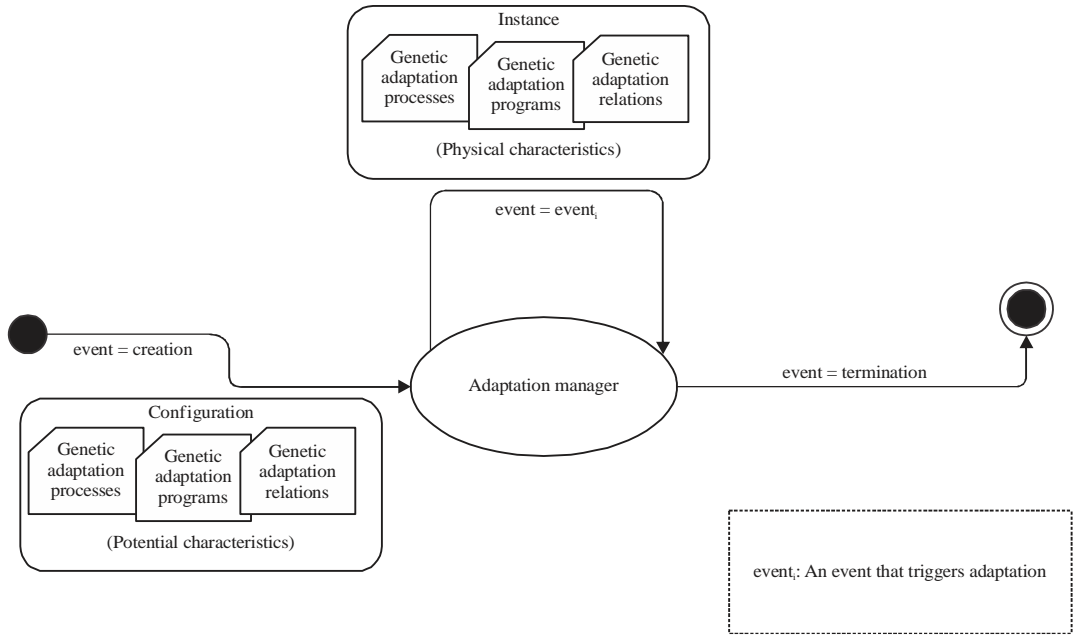

Fig. 1: A genetic adaptation framework

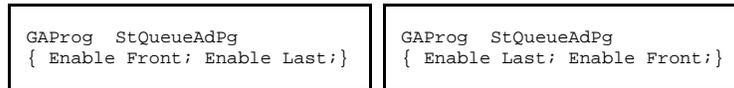

Fig. 2: GAProg of static queue software and static stack software

Where:
- \+      : Repeated once or more
- Enable : A predefined operation that allows a software instance to hold enumerated features (features activation)
- Disable : A predefined operation that allows a software instance to lose enumerated features (features deactivation)

At the initial state (creation of a software instance), all features are disabled. Disabled features are inactive and hence cannot be used until they are enabled.

**Genetic Adaptation Relations (GAR):** To ensure and control the coherence of a software instance, the following dependency relations framework between its features are used.

**Are implied:** These relations ensure the implication between features. This is supported by the following rules:

$$Enable<feature_i>Implies\ Enable<feature_j>$$
//Enabling $feature_i$ will Enable $feature_j$

$$Disable<feature_i>Implies\ Disable<feature_j>$$
//Disabling $feature_i$ will Disable $feature_j$

**Are exclusive:** These relations ensure the exclusion between features. This is supported by the following rules:

$$Enable<feature_i>Excludes\ Enable<feature_j>$$
// Enabling $feature_i$ will Disable $feature_j$

$$Disable<feature_i>Excludes\ Disable<feature_j>$$
//Disabling $feature_i$ will Ensable $feature_j$

**Genetic adaptation process (GAProc):** A genetic adaptation process framework is a process that defines the lifecycle of a software instance, by determining adaptations that are triggered by certain events (Fig. 3). This process is specified as it follows:

GAProc <ID>
{ ((event = $event_i$): $GAProg_i$, $Beh_i$)$^+$ ;}

Where:

event$_i$ : An event that triggers a certain adaptation

GAProg$_i$ : A genetic adaptation program achieving the software instance state$_i$

Beh$_i$ : The adapted-to behavior (Fig. 3)

**Adaptation manager:** The interpretation of a genetic adaptation program is mainly supported by the genetic adaptation relations which ensure the coherence of the adaptation process. The inter/intra relation coherence is ensured at the definition or the update of a software instance. The adaptation manager (which is integrated to software instance) enforces the following rules on its associated software configuration (Fig. 4a) and software instance (Fig. 4b).

**Initial state: Ensuring the coherence of features:**

- **R1:** Each software instance holds, from its software configuration, an initial set of features defined by its genetic adaptation program at its creation. All these features are disabled

- **R2:** Let Enabled_List be the list of the features to be enabled (imposed by an Enable clause in the GAProg)

- **R3:** Let Disabled_List be the list of features to be disabled (imposed by a Disable clause in the GAProg)

- **R4:** The coherence of Enabled_List and Disabled_List is checked: Enabled_List∩Disabled_List = Ø. For each element e of Enabled_List, e should not be implied directly or indirectly by any other element in Disabled_List and e should not be excluded directly or indirectly by any element in Enabled_List. For each element c in Disabled_List, c should not be implied directly or indirectly by any other element in Enabled_List and c should not be excluded directly or indirectly by any element in Disabled_List

**Enable_List and Disable_List processing by scanning genetic adaptation relations:**

- **R5:** The processing of Enabled_List and Disabled_List is carried out by scanning the genetic adaptation relations as it follows:

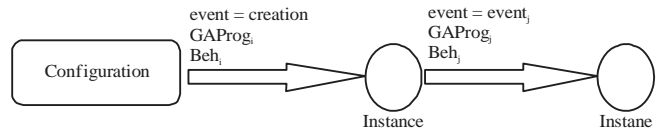

Fig. 3: Genetic adaptation process

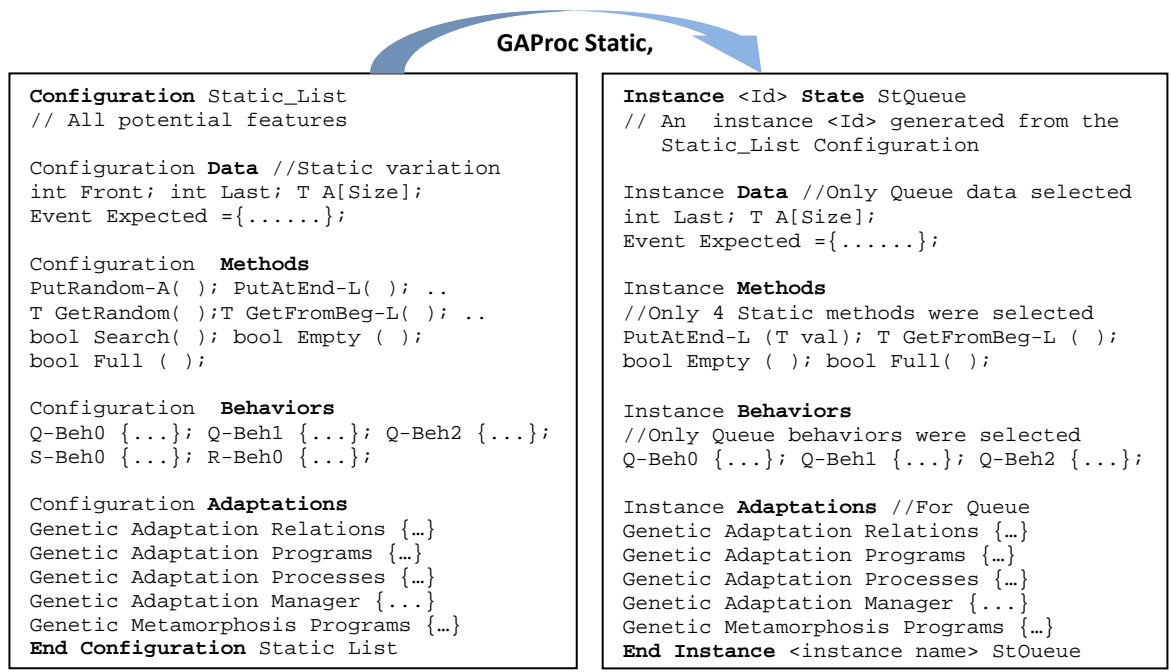

Fig. 4(a-b): (a) A Static_List software configuration and (b) A software instance of Static_ List configuration in StQueue state

- For each element in the Disabled_List, not processed yet: (1) Disable the element, (2) Find all disabled features associated with it. Add them to Disabled_List and (3) Find all enabled features associated with the element. Add them to Enabled_List
- For each element in the Enabled_List, not processed yet: (1) Enable the element, (2) Find all enabled features associated with it. Add them to Enabled_List and (3) Find all disabled features associated with the element. Add them to Disabled_List
- Check for coherence when adding new elements to the two lists (Enabled_List ∩ Disabled_List = ∅)

**Loop on Enabled_List and Disabled_List processing:**

- **R6:** Repeat R4 and R5 until all their elements in the two lists are processed

**Final state:**

- **R7:** Coherence errors cause failure in the adaptation manager. If this process succeeds, Enabled_List will contain features which are enabled, Disabled_List will contain features which are disabled

**Genetic adaptation scope:** In the genetic adaptation process (Fig. 3), a software instance may adapt inside its software configuration or between different software configurations (from one configuration to another).

**Adaptation inside a software configuration:** Inside a software configuration, a software instance may adapt structurally by holding/losing features of its actual software configuration and behaviorally by holding/losing behaviors. Just like natural adaptation, this adaptation is pre-planned in a genetic adaptation process, inherent to a software configuration, defining software instance's lifecycle. At its creation, each software instance holds its own lifecycle (GAProc) that determines the needed adaptation to be achieved genetically and automatically. Once a software instance is created, it holds initial features and genetic information subset of its software configuration; defined explicitly and implicitly by its initial genetic adaptation program (GAProg). Figure 2 shows two GAProgs: StQueue AdPg and StStackAdPg.

Each software instance has a predefined set of events through its lifecycle defined in its GAProc. Each event triggers the software instance to adapt automatically from one state to another. Naturally, the environment may influence this adaptation at any time during the software instance lifecycle. This influence is carried out genetically. While the adaptation by environment (out of scope of this study) affects specific software instances, the genetic adaptation relates to all software instances of the associated software configuration.

**Structural adaptation:** It deals with features (state, data, methods and adapters) adaptations. Figure 5 shows GAProc Static, associated to the software configuration Static_List (Fig. 4a). The software instance <Id> was generated from the software configuration Static_List (first adaptation) at the state StQueue (defined in GAProc Static).

**Behavioral adaptation:** The behavior of a software instance is associated to its features, so the behavioral adaptation is a consequence of features adaptation. A behavior of a software instance is an organization, in the time, of its state enabled features. So, to each state is associated a behavior and thus to

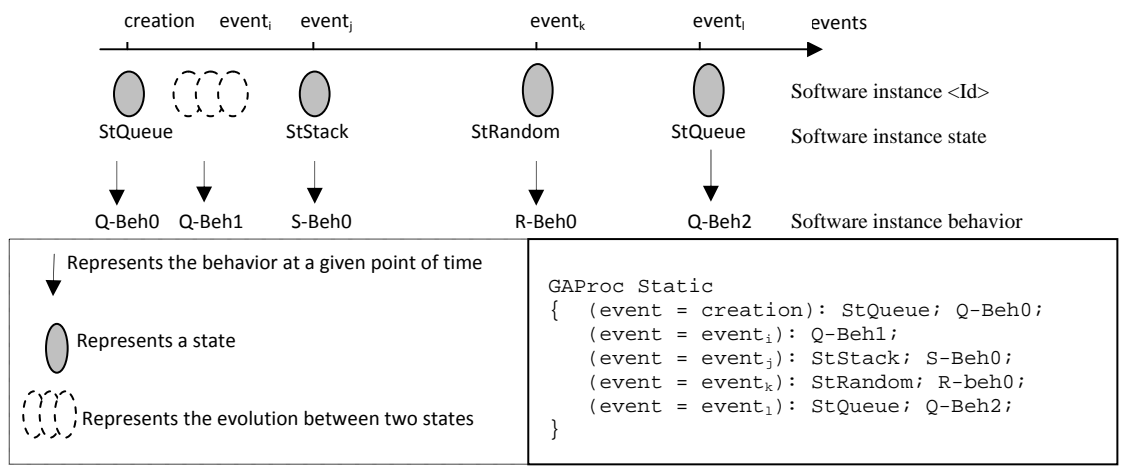

Fig. 5: A genetic lifecycle (GAProc) static of a software instance of Static_List configuration

the structural adaptation is associated a behavioral adaptation (Fig. 6). However, even at the same state, a behavior of an instance may adapt, separating it from the next state. A software instance behavior framework is defined as it follows:

---
Behavior <ID>
{ <feature$_i$> → ((condition)* <feature$_j$>,)$^+$ ; }

---

where, → denotes the right side features will be executed after the left side features if the evaluation of the condition(s) is equivalent to true, + means repeated once or more and * means repeated zero or more.

The behavior Q-Beh0 associated to a software instance of Static_List in a StQueue state, may be defined graphically and textually as it is illustrated in Fig. 6. The behavioral adaptation process is enforced by the following rules:

- **R1:** All involved features must be enabled, at the associated adaptation state, else the process stops
- **R2:** Labeled arrows are conditions on features outputs. Unless these conditions are met, the target features will not be executed

**A running example:** The following application program creates a software instance List1 from the software database List, holding the features specified by the software configuration Static_List (Fig. 4a) and having the lifecycle defined by the GAProc Static (Fig. 5):

---
{ …
List Static_List List1= New Static( );
 // List is the Business software database. Static_List is a software configuration on List, defining all potential features to be held by instances of this software configuration. List1 is a software instance, holding its features from List according to Static_List specification. Static is the GAProc defining the lifecycle of List1 (Fig. 5). List1 will have an initial state StQueue, defined by the GAProg StQueueAdPg (Fig. 2) and will behave according to the behavior Q-Beh0 (Fig. 6).

. . . // List1 is used according to the StQueue state

//Responding to event$_j$; the GAProg StStackAdPg will be activated as illustrated in Fig. 5. It was designed for changing List1 from StQueue state to StStack state (Fig. 7).
… //List 1 is used according to the StStack state
}

---

**Adaptation between software configurations (Metamorphosis):** Between software configurations, a software instance may adapt by losing features and behaviors of its actual software configuration and holding new features and behaviors of another software configuration. The adaptation inside a software configuration was studied; this part deals with the adaptation from one software configuration to another, which is termed by the metamorphosis, i.e., from static queue list to dynamic queue list, from dynamic random list to static random list, etc. A metamorphosis is an adaptation with change (increase, destruction) in features and behaviors, whereas the adaptation inside a software configuration is only in enabling or disabling already held features and behaviors (from the corresponding software configuration). Figure 8 shows a GAProc of a software instance that adapts between two software configurations, Static_List (Fig. 4a) and Dynamic_List (Fig. 9).

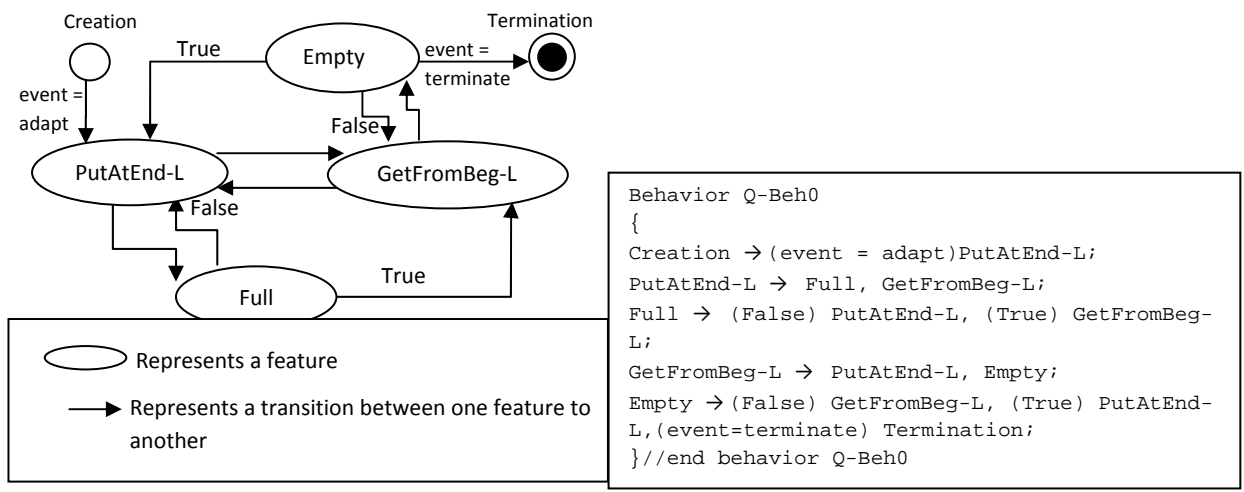

Fig. 6: Behavior Q-Beh0 of a software instance in StQueue state defined textually and graphically

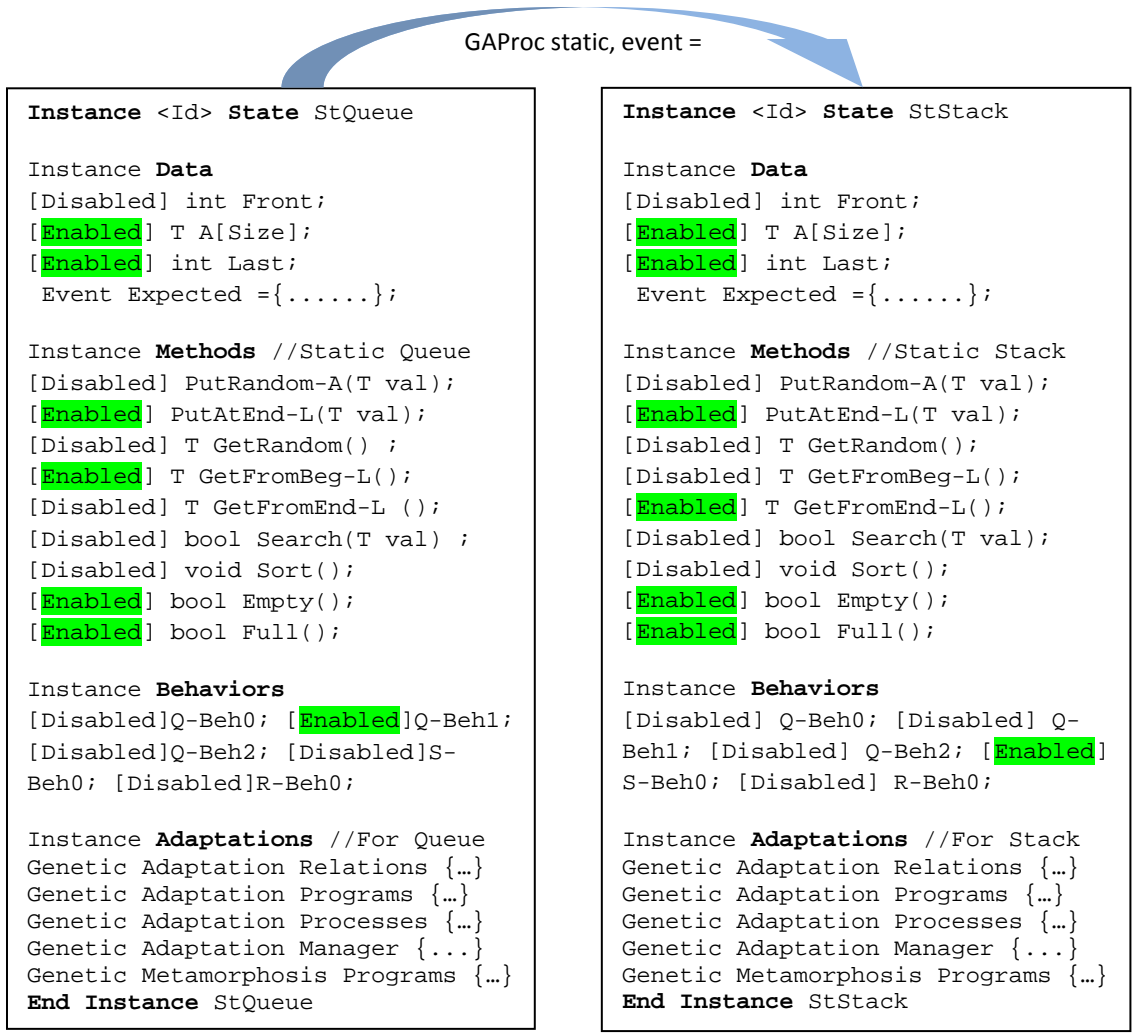

Fig. 7: Software instance adaptation from the state StQueue to the state StStack

```
 GAProc StaticToDynamic
 {
    (event=creation): StQueue, Q-Beh0;
    (event=event_i): StQueueToDyQueue,……
    //StQueueToDyQueue is a Metamorphosis program from StQueue to DyQueue
    (event=event_j): DyStack, Dy-S-Beh0;
    (event=event_k): DyStackToStRandom, R-Beh0;
    //DyStackToStRandom is a Metamorphosis program from DyStack to StRandom
    (event=event_l): StStack, S-Beh0;
 } End GAProc StaticToDynamic
```

Fig. 8: A genetic lifecycle of a software instance adapting between Static_List and Dynamic_List configurations, by metamorphosis programs

The metamorphosis of a software instance SI1, of a software configuration SC1 to a software configuration SC2 is a process which may change SI1 completely (i.e., destruction of old features and holding new ones). It operates like a

```
Configuration  Dynamic_List

Configuration Data
Struct Node { T value; Node *next;} ;
Node *Front; Node *Last;
Event Expected ={......};

Configuration  Methods
//Same as Static_List (Fig. 4) but implemented on Dynamic data

Configuration Behaviors { … }
Configuration Adaptations {…}

End Dynamic List
```

Fig. 9: Dynamic List configuration

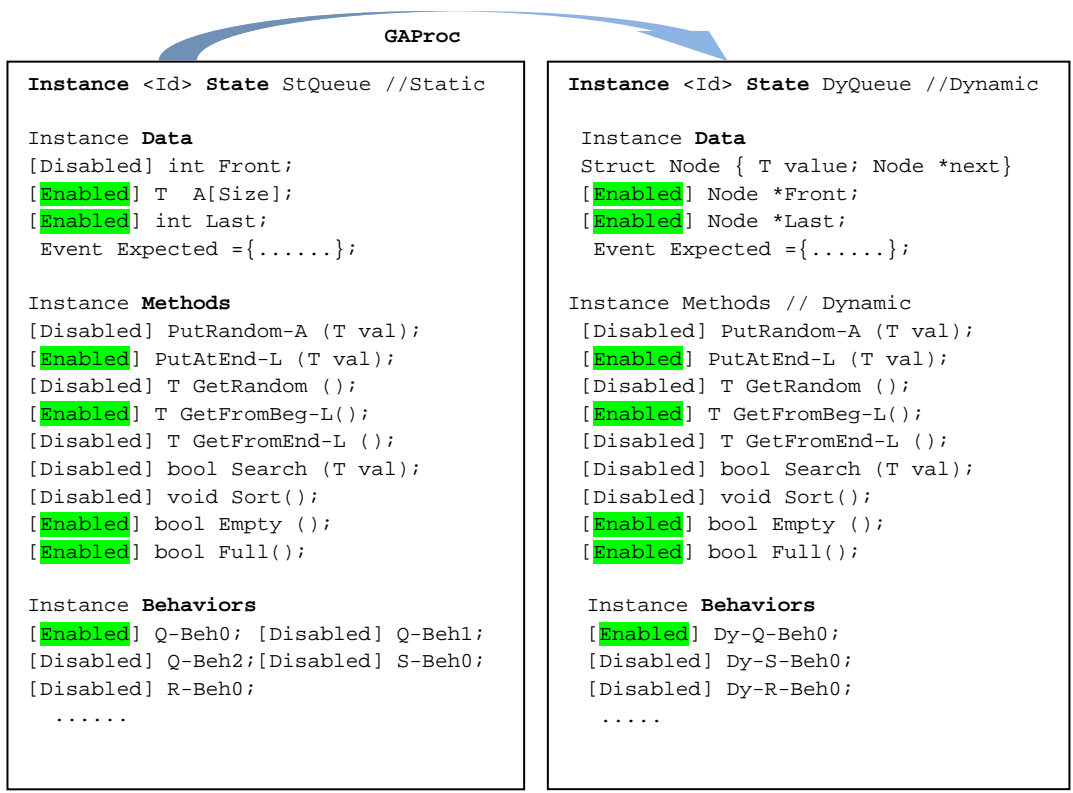

Fig. 10: A software instance adaptation from the state StQueue to the state DyQueue

conversion of SI1 to a new instance SI2 of the software configuration SC2, with a maximum of information transition, such as identity, lifecycle, persistent state information, etc.

A metamorphosis program is defined as it follows:

---

Metamorphosis_Program <ID>
{ Metamorphose to Software configuration <Software configuration_Id>;
 At the Adaptation State <GAProgi> to the Adaptation State <GAProgj>;
 Information transition ensured by the function <Funct_Id>;
}

---

where, Software configuration_Id is the target software configuration, $GAProg_i$ is the current GAProg, $GAProg_j$ is the target GAProg and Funct_Id is the identifier of a user defined function ensuring the transition of specific persistent information from software instance SI1 to software instance SI2.

Figure 10 and 11 show the adaptation of a software instance from StQueue state into a DyQueue state done by the metamorphosis program StQueueToDyQueue.

```
Metamorphosis_Program StQueueToDyQueue
{
   Metamorphose to Configuration Dynamic_List;
   At the Adaptation State StQueue to the Adaptation State DyQueue;
   Information transition ensured by the procedure StQueueToDyQueueTrans;
   procedure StQueueToDyQueueTrans
   {
    While (not Queue.Empty( ))
    {DyQueue.PutAtEnd(StQueue.GetFromBeg ( ) ;}
   }
 }// End Metamorphosis_Program StQueueToDyQueue
```

Fig. 11: Metamorphosis program

**A running example:** The following application program creates an instance List2 from the software database List, holding the features specified by the software configuration Static_List (Fig. 4a) and having the lifecycle defined by the GAProc StaticToDynamic (Fig. 8).

---

{ List Static_List List2= New StaticToDynamic( );
// List is the Business software database. Static_List is a software configuration on List, defining potential features to be held by instances of this software configuration. List2 is a software instance, holding its features from List according to Static_List specification. StaticToDynamic is the GAProc defining the lifecycle of List2 (Fig. 5). List2 will have an initial state StQueue, defined by the GAProg StQueueAdPg (Fig. 2) and will behave according to the behavior Q-Beh0 (Fig. 6)

... // List2 is used according to the StQueue state

//Responding to event$_i$; the GAProc StaticToDynamic will be activated as illustrated in Fig. 8. It was designed for changing List2 from StQueue state to DyQueue state (Fig. 10).
··· //List2 is used according to the Dynamic state
}

---

## RESULTS AND DISCUSSION

Most of the current approaches to self-adaptive systems[1-3] decompose the adaptation process into several processes: Monitoring process which is responsible for collecting data, analyzing or detecting process which is responsible for analyzing the collected data, planning process which responsible for deciding what needs to change and executing process to apply the needed change. In most existing solutions, the adaptation process is assigned to an external adaptation manager that is separate from application logic. An adaptation manager should recognize the four processes (monitoring, analyzing, planning and executing), to control the behavior of self-adaptive software. An important aspect of current self-adaptive systems is feedback loops which control self-adaptation process.

Lots of bio-inspired[8,9] studies, which are better than the conventional ones, have been conducted in the area of self-adaptive software. All of them have focused on the behavior of biological entities (like birds, ants, immunity, etc.) without going in depth into their genetic material that causes this behavior and constitutes the challenge the work presented in this study dealt with. The proposed framework is wholly based on the role of genetic material in the adaptation process. Consequently, it complete the others bio-inspired studies by providing a natural implementing way. The integration of the current bio-inspired approaches (which deal with the natural entities behaviors external modeling) with the proposed framework (which deals with genetics-inspired internal modeling of these behaviors) will lead to homogenous and coherent bio-inspired approaches to self-adaptive software. Without this integration the current bio-inspiration remains hybrid of natural inspiration and artificial (with computer paradigms) representation.

In contrast to self-adaptive systems built using control engineering concepts (current artificial approaches), self-adaptive systems in nature, as in the proposed framework, do not often have a single clearly visible control loop. There is no separation between the application, the adaptation process, controller and the other elements presented in advanced control schemes. Further, the systems are highly decentralized in such a way that the software have no sense of the global goal but rather it is the interaction of their local behaviors.

A first evaluation was by reusing this framework in software object evolution. The conceptual and practical obtained results were valuable in object-oriented paradigm enrichment. However, this separate evaluation is not sufficient. The performance evaluation of the actual bio-inspired hybrid approaches against the proposed integrated ones must be evaluated. Unfortunately, it is impossible to achieve currently this evaluation because the integrated approaches are just what this study proposes but it might be expected that the integrated ones will be better (in the whole self-adaptive software engineering processes) than the hybrid current ones. The homogeneity of approaches has its important impact.

## CONCLUSION

This study proposes a solution to the limitation of the current approaches (in bio-inspired self-adaptive software modeling) to external behaviors of biological entities without taking into account the genetic material supporting them. This solution consists of a genetic framework model for self-adaptive software; inspired by the adaptation (evolution) of biological entities and taking into consideration the role of genes in the adaptation process. The obtained framework model completes the others bio-inspired studies by providing a natural implementing way. The integration of the current bio-inspired approaches (which deal with the natural entities behaviors external modeling) with the proposed framework (which deals with genetics-inspired internal modeling of these behaviors) will lead to homogenous and coherent bio-inspired approaches to self-adaptive software. The proposed framework is limited to self-adaptations predicted at the requirements and design steps in self-adaptive software engineering, which is significant in practice. However, the unpredicted adaptation (to unpredicted errors, environment requirements, etc.) will be a genetics-inspired approach real challenge.

## SIGNIFICANCE STATEMENTS

- All of the current bio-inspired self-adaptive software studies deal with natural entities external behaviors modeling without taking into account their genetic material
- The proposed genetic framework model for self-adaptive software deals with taking into account this genetic material for providing internal model of natural entities behaviors.
- The proposed model leads to complete the bio-inspired self-adaptive software modeling approaches by supporting entities behaviors external models as well as entities behaviors internal models
- This leads to homogenous and fully bio-inspired approaches to self-adaptive software modeling

## ACKNOWLEDGMENT

This study is sponsored from 12/12/2013 to 12/12/2016 by Philadelphia University, through the research project "Bio-inspired Systems Variability Modeling" at the Research Laboratory on Bio-inspired Systems.